\documentclass[prl,twocolumn,showpacs,preprintnumbers,amsmath,amssymb]{revtex4}
\usepackage{amsfonts}
\usepackage{graphicx}
\usepackage{dcolumn}
\usepackage{float}
\usepackage{bm}

\newcommand{\com}[1]{}

\begin{document}

\title{Kinetics of the Phase Separation Transition in Cold-Atom Boson-Fermion Mixtures}

\author{Dmitry Solenov\footnote{E-mail: solenov@clarkson.edu}
and Dmitry Mozyrsky\footnote{E-mail: mozyrsky@lanl.gov}}
\affiliation{$^*$Department of Physics, Clarkson University,
Potsdam, New York 13699-5820, USA \\
$^\dag$Theoretical Division (T-4), Los Alamos National Laboratory,
Los Alamos, NM 87545, USA}


\begin{abstract}
We study the kinetics of the first order phase separation
transition in boson-fermion cold-atom mixtures. At sufficiently
low temperatures such a transition is driven by quantum
fluctuations responsible for the formation of critical nuclei of a
stable phase. Based on a microscopic description of interacting
boson-fermion mixtures we derive an effective action for the
critical droplet and obtain an asymptotic expression for the
nucleation rate in the vicinity of the phase transition and near
the spinodal instability of the mixed phase. We also discuss
effects of dissipation which play a dominant role close to the
transition point, and identify the regimes where quantum
nucleation can be experimentally observed in cold-atom systems.

\end{abstract}


\maketitle

Kinetics of the first order phase transitions at ultralow
temperatures has received considerable attention in connection
with several problems ranging from the decay of false vacuum in
field theoretical models of the early Universe~\cite{Coleman} to
the phase separation in $^3$He-$^4$He
mixtures~\cite{LifshitzKhokhlov}. In the latter case it has been
argued that below a certain temperature (of the order of a few
tens of mK) formation of pure $^3$He phase from a supersaturated
mixture occurs via the process of quantum nucleation, where
critical nuclei overcome the potential barrier (due to the surface
energy between the two phases) by means of quantum tunneling. It
has been predicted theoretically~\cite{LifshitzKagan} that the
rate of such nucleation behaves as $\exp{(-C/\Delta\mu^{7/2})}$,
where $\Delta\mu$ is the degree of supersaturation, i.e., the
difference in chemical potentials of the two phases, and $C$ is
related to the coefficients of the phenomenological
Ginzburg-Landau expansion of the free energy near the point of the
phase transition. Measurements carried out by several
groups~\cite{HeExperiment} seem to confirm that at sufficiently
low temperatures the kinetics of the phase separation in
$^3$He-$^4$He becomes temperature independent; however, they have
been unable to verify the expected dependence of the nucleation
rate on the systems' parameters (i.e., $\Delta\mu$, etc) - partly
due to the poor knowledge of microscopic interactions between
particles in such a strongly correlated system.

We argue that contemporary cold atom systems provide a perfect
setup for studying and observing the kinetics of such a phase
separation transition. Atomic mixtures, such as boson-fermion
mixture, are commonly realized in experiments on sympathetic
cooling, where one of the species (typically bosons) plays the
role of a coolant~\cite{BosCool}. Another interesting realization
of boson-fermion mixture has been demonstrated in a two-component
fermion system, where strongly bound Cooper pairs correspond to
bosons interacting with unpaired fermion atoms~\cite{Ketterle}.
Starting from a microscopic description of a boson-fermion mixture
we derive an effective action for the order parameter (the
condensate density) taking into account fermion-boson interaction.
We show explicitly that the classical potential for the order
parameter due to such interaction has two minima corresponding to
the two phases of the system (mixed and phase separated). The two
minima are separated by the finite energy barrier, which points
out that such a transition is indeed of the first
order~\cite{bosons}. We then derive an expression for the
nucleation (tunneling) rate of the critical droplet of the pure
fermion phase near the phase transition line and near the line of
absolute (spinodal) instability  of the mixed phase. We also
evaluate the role of dissipation~\cite{CaldeiraLeggett} in the
quantum nucleation process and find that near the line of the
first order phase transition it changes the leading asymptotic
behavior of the nucleation rate on the degree of supersaturation.

We consider a Bose-Einstein condensate interacting with a single
species of fermions (in the same spin state). Interactions in such
a mixture are characterized by two scattering lengths $a_{BB}$ and
$a_{BF}$. Fermions and bosons interact through contact potential
$\lambda_{BF}\delta(\mathbf{r}-\mathbf{r}')$, contributing term
$\lambda_{BF}\rho \psi_F^\dag\psi_F$, where $\lambda_{BF} =
2\pi\hbar^2a_{BF}(1/m_B + 1/m_F)$; $\sqrt{\rho} e^{i\phi}$ and
$\psi_F$ are bosonic and fermionic fields respectively. In
addition, boson-boson interaction gives rise to another term
$\lambda_{BB}\rho^2/2$ in the Hamiltonian density, with
$\lambda_{BB}=4\pi\hbar^2a_{BB}/m_B$. The direct coupling between
fermions is negligible due to the exclusion principle in the
s-scattering channel (p-wave scattering is usually small compared
to s-wave fermion-boson and boson-boson interactions). For the
purposes of present calculation we can neglect the spatial
dependence of the trapping potential and assume that the local
densities of fermions and bosons are set by the chemical
potentials $\mu_F$ and $\mu_B$. Indeed, since the first order
transition occurs at finite coherence length (to be defined
below), the shape of the trap potential should play little role in
the dynamics of the phase transition as long as the effective size
of the trap is much greater than the coherence length.

\begin{figure}
\includegraphics[width=7.5cm]{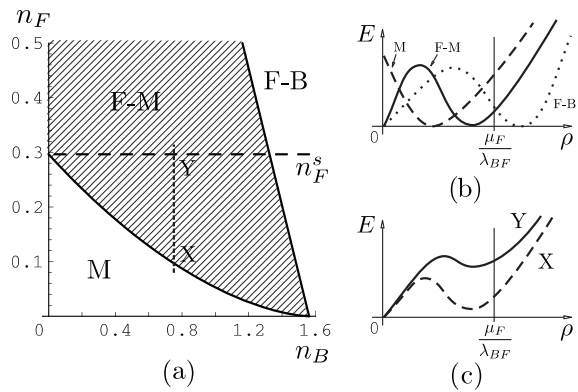}
\caption{(a) The phase diagram of the equilibrium boson-fermionic
mixture. The bosonic density $n_B$ is in units of $\frac{9\pi}{4}
({a_{BB}^2}/{a_{BF}^5})({m_F^2}/{m_B^2})
\left({m_F}/{m_B}+1\right)^{-5}$ and the fermionic density $n_F$
is in units of $\frac{9\pi}{2}
({a_{BB}^3}/{a_{BF}^6})({m_F^3}/{m_B^3})$ $\times
\left({m_F}/{m_B}+1\right)^{-6}$. The area where pure fermionic
fraction can coexist with the mixture (F-M) is hatched. The
mixture is unstable above the dashed line. (b) Three possible
curves for $E(\rho)$ at equilibrium (sketch). (c) The sketch of
$E(\rho)$ for out-of-equilibrium mixture: above the lower phase
separation curve (dashed curve) and below the instability line
(solid curve). The corresponding points are shown on the phase
diagram (a).}\label{fig1.eps}
\end{figure}

It is convenient to describe the system in terms of the bosonic
field only, averaging $e^{-H/k_BT}$ ($H$ is the overall
Hamiltonian) with respect to the fermionic field. Such averaging
can be easily carried out within mean field, i.e., the
Thomas-Fermi approximation~\cite{Viverit,Mozyrsky}, so that the
calculation reduces to the evaluation of the canonical partition
function (or free energy) of the free fermions with effective
chemical potential $\mu_F-\lambda_{BF}\rho$. In the
zero-temperature limit the effective classical potential density
for the bosons is
\begin{eqnarray}\label{Eq:Eeff}
E(\rho)\!\! =  &-&\!\! \mu_B \rho  + \frac{1}{2}\lambda_{BB}
\rho^2
\\ \nonumber
&-&\!\! \frac{(2m_F)^{3/2}}{15\pi^2\hbar^3} \left(\mu_F  -
\lambda_{BF} \rho\right)^{5/2} \theta(\mu_F -
\lambda_{BF}\rho),
\end{eqnarray}
where $\theta(x)$ is a step function. It is known that a
fermion-boson mixture with pointlike interactions exhibits three
different phases \cite{Viverit}: uniform mixture of bosons and
fermions (M), pure fermion fraction coexisting with the mixture
(F-M), and separated bosonic and fermionic fractions (F-B), see
Fig~\ref{fig1.eps}a. Equation~(\ref{Eq:Eeff}) is sufficient to
obtain the entire structure of the phase diagram. Indeed, one can
notice that $E(\rho)$ has either one or two local minima separated
by the barrier (see Fig.~\ref{fig1.eps}b). The one at $\rho=0$
corresponds to pure fermionic phase. The other, $\rho=\rho_0$,
characterizes the mixture if $\rho_0<\mu_F/\lambda_{BF}$, or pure
bosons, if $\rho_0\ge \mu_F/\lambda_{BF}$. At low densities only
the mixture exists at equilibrium. For higher densities the
mixture becomes metastable and, eventually, unstable above the
absolute (spinodal) instability line. The phase transition line
between M and F-M phases can be obtained from the condition
$E(0)=E(\rho_0)$ in Eq.~(\ref{Eq:Eeff}) in the parametric form:
$n_F^0=A^{-3}y^3/8$, $n_B^0=A^{-2}[1-y^2]/4$, where the densities
are in the units of Fig.~\ref{fig1.eps}a, $A =
3\pi^2\hbar^3\lambda_{BB}/(2m_F)^{3/2}\lambda_{BF}^2 \mu
_F^{1/2}$, and parameter $y$ is obtained from the equation $2 + 4y
+ 6y^2 + 3y^3 - 5A(1 + y)^2 = 0$. Note that the solution with
$0<y<1$ exists only for $2/5\leq A \leq 3/4$.

In what follows we consider the system with fixed global density
of the bosons $n_B$. Upon variation of fermion density in the
metastable region, i.e., between points X and Y in
Fig.~\ref{fig1.eps}a, the classical potential $E(\rho)$ varies
continuously between the two situations shown schematically in
Fig.~\ref{fig1.eps}c: with nearly equal minima in the vicinity of
the phase transition point (at $n_F=n_F^0+\Delta n_F$) and
vanishingly small barrier near the absolute instability at
$n_F=n_F^s=8/27$ (in dimensionless units). In the rest of this
Letter we study kinetics of the system, i.e., the rate of
formation of the stable pure fermion phase out of the metastable
mixed phase in these two limiting cases. To do so we consider an
effective Lagrangian density of the bosons, which can be written
in density-phase variables as
\begin{eqnarray}\label{Eq:lag}
L= \hbar\rho{\dot\phi}+\frac{\hbar^2} {2m_B}\rho(\nabla \phi)^2+\frac{\hbar^2}
{8m_B\rho}(\nabla\rho)^2 +E(\rho) .
\end{eqnarray}
The first term in Eq.~(\ref{Eq:lag}) is the Berry phase term,
while the second and the third terms are the kinetic energy of the
superfluid. The Thomas-Fermi approximation utilized in
Eqs.~(\ref{Eq:Eeff},\ref{Eq:lag}) implies that renormalization of
the boson kinetic energy arising due to the non-locality of the
fermionic response function is relatively small. A straightforward
perturbative estimate (to the second order in $\lambda_{BF}$)
yields the gradient term, i.e. the correction to the Thomas-Fermi,
$\sim (m_F^{3/2}\!\lambda_{BF}^2 \mu_F^{1/2} /\hbar^3
k_F^2)(\nabla\rho)^2$. Comparing this term with the third term in
Eq.~(\ref{Eq:lag}) we see that near the phase transition line it
is smaller by factor $\sim(k_F l)^{-2}$, where $l$ is boson
coherence length, $l\sim a_{BB}/g_B$, and $g_B^2=a_{BB}^3\rho_0$
is conventional boson gas parameter, which, in terms of the
dimensionless $n_B$ (as in Fig.1) is
\begin{equation}\nonumber
g_B^2 = {9\pi\over 4}\left({a_{BB}\over
a_{BF}}\right)^5\left({m_F\over m_B}\right)^2\left({m_F\over
m_B}+1\right)^{-5}n_B.
\end{equation}
For $g_B\lesssim 0.1$ and not too small $n_F$ (e.g. for $n_B\sim
0.4$), $(k_Fl)^2\sim 50$, and thus the Thomas-Fermi approximation is well
justified.

The decay rate per unit volume ($\Gamma/V$) from a metastable
state (at $\rho_0$) can be obtained by calculating the classical
action for the transition between states with $\rho=\rho_0$ and
$\rho=0$ in the imaginary time formalism by following the
prescription of Ref.~\cite{Coleman}. Namely, $\Gamma/V
\sim\exp(-S/\hbar)$, where the action $S= \int
dtd\mathbf{r}L(\rho,\phi)$ is evaluated over the classical
(extremal) trajectory, defined by equations $\delta S/\delta \phi
=0$ and $\delta S/\delta \rho =0$. The first of these equations is
a continuity equation, $\partial _t \rho  + \nabla \left( {\rho
{\bf{u}}} \right) = 0$ with $\mathbf{u} = \hbar\nabla\phi/m_B$. It
can be easily solved in the spherically symmetric case yielding
$\mathbf{u} = ({\mathbf{\hat{r}}}/{r^2\rho})\int_0^r
dr^\prime{{r^\prime}^2\partial _t \rho}$.  In terms of single
variable $\rho$ the action can be cast in the form
\begin{equation}\label{Eq:S0}
S\!\!=\!4\pi\!\!\!\int\!\! dtdrr^2\!\!\left[\!\frac{m_B}{2\rho}\!
\left(\!\!\frac{1}{\,r^2\!\!}\!\int_0^r\!\!\!\!\!
dr'r'^2\partial_t\rho\!\right)^{\!\!2}\!\!\!
+\!\frac{\hbar^2(\nabla\rho)^2}{8m_B\rho}\!+\!
E(\rho)\!\right]\!\!.\!\!
\end{equation}

Evaluation of the extremum of $S$ in the two cases of interest,
i.e., near the line of the phase transition (at point X in
Fig.~\ref{fig1.eps}a) and near the absolute instability line (at
point Y), has been formally carried out in
Ref.~\cite{LifshitzKagan}. In the former case the solution can be
parameterized as $\rho(r,t) \approx \rho_0 \theta[r-R(t)]$ (the
thin wall approximation), where $R(t)$ is the the radius of the
critical droplet and $\rho_0$ is bosonic density of the mixed
phase as before. Within such an approximation the action $S$ can
be formulated in terms of $R(t)$ as
\begin{equation}\label{Eq:SR}
S_X\!\!= 4\pi \!\!\! \int\!\! dt\left[\frac{m_B\rho_0}{2}R^3
\left(\partial_tR\right)^2 + \sigma R^2 - \frac{\rho_0
\Delta\mu}{3} R^3 \right],
\end{equation}
where the effective surface tension coefficient is $\sigma
=\sqrt{{\hbar^2}/{2m_B}} \int d\sqrt{\rho}\sqrt{E(\rho)}$ and for
$\Delta n_F/n_F^0\ll 1$ we have $\Delta\mu = \lambda_{BF}K\Delta
n_F$, with $K\!\!\!=\![{\frac{2}{3}(n_F^{0\,2/3}\!\!\!+
\!n_B)^{3/2}\!\!-\!n_Bn_F^{0\,1/3}\!\!-\!n_F^0}]\!/{n_Bn_F^{0\,1/3}}\!\!$,
where $n_B$ and $n_F^0$ are dimensionless as in
Fig.~\ref{fig1.eps}a ($n_B$ varies between $0$ and $25/16$).
Evaluating the surface tension coefficient according to
Eq.~(\ref{Eq:Eeff}) and extremizing the action $S_X$, one
obtains~\cite{detailes}
\begin{equation}\label{Eq:R1}
\ln{\Gamma_X/\Gamma_0}=-0.0056 \frac{n_B^{11/2}}{g_B
K^{7/2}}\left({n_F^s\over \Delta n_F}\right)^{7/2}.
\end{equation}
Precise evaluation of coefficient $\Gamma_0$ lies out of the scope
of present calculations. It can be estimated, however, as
$\Gamma_0/V\sim \omega_0/l^3$, where $\omega_0$ is an ``attempt''
frequency. From the uncertainty principle $\omega_0\sim \hbar/2m_B
l^2$ and thus $\Gamma_0/V\sim \hbar/m_Bl^5$.

As expected the tunneling exponent, i.e. the right-hand side of
Eq.~(\ref{Eq:R1}), is singular in the degree of metastability
$\Delta n_F$ and diverges as $\Delta n_F^{-7/2}$
\cite{LifshitzKagan}. Equation (\ref{Eq:R1}) also indicates that
the rate of nucleation is exponentially small in the dilute limit,
i.e., for $g_B\ll 1$. Note that appearance of the dimensionless
boson density $n_B$ in the numerator in Eq.~(\ref{Eq:R1}) does not
fix this problem, since the thin wall approximation (nucleation)
requires sufficiently high energy barrier, e.g.
Fig.~\ref{fig1.eps}(b,c), which is not the case when $n_B\ll 1$.
However, due to the smallness of the numerical coefficient in the
right-hand side of Eq.~(\ref{Eq:R1}) one can hope that quantum
nucleation is observable in sufficiently strongly coupled systems
(which are presently realizable with the use of Feshbach
resonance). Indeed, for $g_B\sim 0.1$, $n_B\sim 0.4$, and $\Delta
n_F/n_F^s = 0.15$, the coefficient $K \sim 0.27$ and the tunneling
exponent is $\sim -27$. For the same parameters and $a_{BB}\sim
20$ a.u., $\Gamma_0/V$ can be estimated to be $\sim
10^{11}$$s^{-1}\mu m^{-3}$, which yields nucleation rate
$\Gamma_X/V\sim 1$~$s^{-1}\mu m^{-3}$.

Let us now evaluate the rate of transitions near the absolute
instability of the mixed phase (at point Y in the phase diagram in
Fig.~\ref{fig1.eps}a). Since the energy barrier in $E(\rho)$ is
now relatively small and $E(\rho_0)\gg E(0)$, see
Fig.~\ref{fig1.eps}c, one should pay attention only to the
vicinity of the metastable minimum at $\rho_0$. Expanding action
in Eq.~(\ref{Eq:S0}) in $\delta\rho$, $\rho = \rho_0+\delta\rho$,
and retaining the second and the third order terms (in
$\delta\rho$) in potential energy $E(\rho)$ and only the second
order terms in the kinetic energy (the third order kinetic terms
contain gradients and thus are small compared to the the third
order potential terms in the limit of diverging coherence length
near the point of absolute instability) one obtains
\begin{eqnarray}\label{Eq:SY}
S_Y   = 4\pi\int dtdrr^2 \left[\frac{m_B}{2\rho_0}
\left(\frac{1}{r^2} \int_0^r dr^\prime
{r^\prime}^2\partial_t\delta\rho\right)^2 \right.
\\ \nonumber
\left.+\frac{\hbar^2}{8m_B}\frac{\left(\nabla\delta\rho\right)^2}{\rho_0}
+ a\delta\rho^2 +b\delta\rho^3\right],
\end{eqnarray}
where
\begin{equation}\nonumber
a = {2\pi\hbar^2 a_{BB}\over {3m_B}}\left( 1-{n_F\over
n_F^s}\right); \ b={\hbar^2a_{BF}^5m_B\over
3a_{BB}m_F^2}\left(1+{m_F\over m_B}\right)^5\,.
\end{equation}
The extremum of $S_Y$ is difficult to evaluate exactly. However,
by introducing dimensionless variables $x=r\sqrt{8m_B\rho_0
a}/\hbar$, $\tau = 4\rho_0 a t/\hbar$, and $p=\delta\rho b/a$, the
action  $S_Y$ can be rewritten as $const\times s_Y$, where
$s_Y[p(x,\tau)]$ is a parameter-independent functional, the
extremum of which is a c-number. Its value can be estimated by
variational ansatz $p=-p_0\exp{(-\alpha x^2 - \beta \tau^2)}$,
where $\alpha$, $\beta$ and $p_0$ are variational parameters. Upon
a straightforward calculation one obtains
\begin{equation}\label{Eq:R2}
\ln{\Gamma_Y/\Gamma_0}=-{0.324\over g_B n_B^2}\left(1-{n_F\over
n_F^s}\right)^{1/2},
\end{equation}
Again we see that tunneling exponent is controlled by the inverse
boson gas parameter $g_B$. The exponent vanishes when fermion
density $n_F$ reaches value $n_F^s$, where the effective energy
barrier disappears, see Fig.~\ref{fig1.eps}c. In this respect
Eq.~(\ref{Eq:R2}) is similar to results on macroscopic quantum
tunneling (MQT) in systems of trapped bosons with attractive
interactions~\cite{UedaLeggett}, but with an important
distinction: unlike in the latter case, the height of the
potential barrier and thus the tunneling exponent for the critical
droplet are controlled not by the total number of particles in the
trap but the local densities. Therefore, for sufficiently large
numbers of particles in the trap the tunneling exponent can be
fine-tuned with a desired accuracy, which makes it possible to
observe the MQT rate in a well controlled and predictable regime.

We now consider effects of dissipation on the kinetics of the
first order phase transitions. The Thomas-Fermi approximation
utilized in the derivation of the effective potential $E(\rho)$
implies that fermions instantaneously adjust to the local
variation of boson density. Therefore it is instructive to proceed
beyond this approximation and evaluate the effect of the
excitation of particle-hole pairs in the fermionic subsystem on
the dynamics of the transition. This effect can be estimated by
considering the second order frequency dependent correction (in boson density $\rho$)
produced by the interaction $\lambda_{BF}\psi_F^\dag\psi_F\rho$.
The resulting contribution to the effective action can be cast in
the form
\begin{equation}\label{Eq:DEC-vac-pol}
\Delta S=\frac{{\lambda _{BF}^2 }}{{2\hbar }} \!\!\int\!\!
{\frac{{d{\bf{q}}}}{{\left( {2\pi } \right)^3 }}\frac{{d\omega
}}{{2\pi }}\left(\frac{{m_F^2 \left| \omega  \right|}}{{4\pi \hbar
^2 q}} +... \right)\left| {\rho (q,\omega )} \right|^2 },
\end{equation}
where the $\omega$ and $\bf q$ independent term (omitted above)
have already been included in Eq.~(\ref{Eq:Eeff}), as can be
verified by expanding $E(\rho)$ to the order $O(\rho^2)$. Again we
consider the two cases: near the phase transition and near the
absolute instability. Utilizing the thin wall approximation,
$\rho(r,t) \approx \rho_0 \theta[r-R(t)]$, we obtain that near the
line of the phase transition the action in
Eq.~(\ref{Eq:DEC-vac-pol}) can be written in terms of the radius
of the critical droplet as
\begin{eqnarray}\label{Eq:Sdec}
\Delta S_X=\frac{\lambda_{BF}^2 m_F^2 \rho_0^2}{16\pi^2\hbar^3}\
{\cal P}\int {dt dt'\over (t-t')^2}\left\{R(t)^3R(t')
\phantom{\frac{R}{R}} \right.
\\ \nonumber
\left. + R(t)R(t')^3 + \frac{1}{2}[R(t)^2 - R(t')^2]^2
\ln\left|\frac{R(t)-R(t')} {R(t)+R(t')}\right|\right\}.
\end{eqnarray}
The first two terms in the right-hand side of Eq.~(\ref{Eq:Sdec})
arise due to the restructuring of the fermionic density of states
inside the droplet in the course of its expansion, while the last
term can be viewed as coupling between droplet's surface and
particle-hole excitation in Fermi sea. Exact evaluation of
extremal action (\ref{Eq:SR}) with the correction (\ref{Eq:Sdec})
is difficult and therefore we use the variational technique to
evaluate the tunneling exponent. A natural ansatz is $R(t)=R_0
e^{-\alpha t^2}$, where the turning point $R_0$ and coefficient
$\alpha$ are variational parameters . We find that correction due
to $\Delta S_X$ strongly alters the tunneling exponent in the
region $K\Delta n_F/n_F^s\!\!\ll\!\eta_{dis}\!=
g_B^{2/5}n_B^{9/5}(m_F\!/m_B)^{8/5}$:
\begin{equation}\label{Eq:lnGmDEC}
\ln\Gamma'_X/\Gamma_0 \approx - 0.04
\frac{n_B^{32/5}}{g_B^{4/5}K^4}\left({m_F\over
m_B}\right)^{4/5}\left(n_F^s\over\Delta n_F\right)^4,
\end{equation}
while in the opposite (nondissipative) limit $\Delta n_F/n_F^s
\gg\eta_{dis}$ the tunneling exponent is given by
Eq.~(\ref{Eq:R1}). We see from Eq.~(\ref{Eq:lnGmDEC}) that the
influence of dissipation on the dynamics of nucleation is
significant for $m_B\leq m_F$.\com{: indeed for $m_F\!\sim\!m_B$,
even in the dense limit ($g_B\!\sim\!1$ and $n_B\!\sim\!1$), for
$n_F^s/\Delta n_F = 10$  we obtain that $\ln\Gamma'_X/\Gamma_0
\sim -400$, and thus $\Gamma^\prime_X\sim 10^{-200}\, s^{-1}\mu
m^{-3}$. Thus we conclude that dynamics of quantum nucleation near
the phase transition line can be systematically observable only
for cold atom systems with sufficiently small mass ratio
$m_F/m_B$.}

Near the absolute instability (for $n_F\!\!\!\to\!\! n_F^s$) the
contribution of dissipation, e.g. Eq.~(\ref{Eq:DEC-vac-pol}), can
be estimated by using the same variational ansatz as in derivation
of Eq.~(\ref{Eq:R2}). We find that the tunneling exponent acquires
an additional term $\sim -(m_F g_B n_B^2/m_B)^{4/5}$ (the
numerical coefficient in Eq.~(\ref{Eq:R2}) does not change
significantly when $\Delta S$ is taken into account). This term is
independent of fermion density and is again controlled by the mass
ratio $m_F/m_B$. Therefore for not too high fermion/boson mass
ratio, it is of the order $g_B^{4/5}$ and thus dissipation does
not significantly alter the dynamics of the phase transition (MQT)
in the absolute instability region.

Finally we estimate the crossover temperature between thermal and
quantum regimes for the transition dynamics in the two cases of
interest. The thermal regime becomes effective when the energy gap
between the exited and the ground state energies of the metastable
mixture is of the order $k_BT$ \cite{Gorokhov}. This energy
difference is $\hbar\Gamma_0\!\!\!\sim\!\!\hbar^2\!\!/2m_B l^2\!$.
Near the phase transition curve it gives the relation
$g_B^{2/3}\!n_B\!\!\sim\!T\!/T_c$, where $T_c$ is the
Bose-Einstein condensate transition temperature. Since we assume
$T\!\!\!\ll\!T_c$ and $n_B\!\sim\!1$, the tunneling is clearly the
dominant process of nucleation. Near the spinodal instability we
obtain $g_B^{2/3}(1\!-\!n_F/n_F^s)\!\sim\!T/T_c$. In this case, as
expected, thermal excitation energy bounds the coherence length
$l\!\sim\!(1\!-\!n_F/n_F^s)^{-1/2}$ for which the transition is
dominated by MQT mechanism.

\com{ In summary we have studied kinetics of the phase separation
transition in boson-fermion cold atom mixtures. We have evaluated
the transition rates in the vicinity of the phase transition line
and near the absolute instability of the mixed phase. We have also
studied the influence of dissipation on the dynamics for such the
transition. We argue that under certain conditions the dynamics of
quantum nucleation can be observed in contemporary cold atom
systems.}

We thank Eddy Timmermans for valuable discussions and comments.
The work is supported by the US DOE.


\end{document}